\documentclass[aip, superscriptaddress, preprint]{revtex4}
\usepackage{amsmath, amssymb}
\usepackage{graphicx}
\usepackage{bm}
\usepackage{siunitx} 
\usepackage{hyperref}
\hypersetup{
            colorlinks=true,
            linkcolor=blue,
            anchorcolor=blue,
            citecolor=blue
            }
\newcommand{\Fig}[1]{Fig. \ref{#1}}
\usepackage{changes}
\begin{document}
\title{Dual function spin-wave logic gates based on electric field control magnetic anisotropy boundary}

\author{Kang Wang}
\affiliation{Center for Spintronics and Quantum Systems, State Key Laboratory for Mechanical Behavior of Materials, School of Materials Science and Engineering, Xi'an Jiaotong University, Xi'an, Shaanxi, 710049, China}

\author{Shaojie Hu}
\email[]{shaojiehu@mail.xjtu.edu.cn}
\affiliation{Center for Spintronics and Quantum Systems, State Key Laboratory for Mechanical Behavior of Materials, School of Materials Science and Engineering, Xi'an Jiaotong University, Xi'an, Shaanxi, 710049, China}

\author{Fupeng Gao}
\affiliation{Center for Spintronics and Quantum Systems, State Key Laboratory for Mechanical Behavior of Materials, School of Materials Science and Engineering, Xi'an Jiaotong University, Xi'an, Shaanxi, 710049, China}

\author{Miaoxin Wang}
\affiliation{Center for Spintronics and Quantum Systems, State Key Laboratory for Mechanical Behavior of Materials, School of Materials Science and Engineering, Xi'an Jiaotong University, Xi'an, Shaanxi, 710049, China}

\author{Dawei Wang}
\affiliation{School of Microelectronics $\&$ State Key Laboratory for Mechanical Behavior of Materials, Xi'an Jiaotong University, Xi'an 710049, China
}


\date{\today}

\begin{abstract}
Spin waves (SWs) have been considered a promising candidate for encoding information with lower power consumption. Here,
we propose the dual function SW logic gates based on the electric field controlling the SW propagation in the Fe film of $\rm Fe/BaTiO_3$ heterostructure with the motion of magnetic anisotropy boundary (MAB). 
We show micromagnetic simulations to validate the AND-OR and NAND-NOR logic gates. Our research may find a path for simplifying integrated logic circuits using such dual function SW logic gates.
\end{abstract}
\maketitle

With the rapid development in the miniaturization of electronic devices, power consumption has become one important issue because of the unbearable Joule heating.\cite{2007Vogel, 2017Theis}
To overcome this issue, seeking new charge-neutral information carriers has spurred great attention. One representative charge-neutral information carrier is spin angular momentum, the electron’s another degree of freedom. Spin waves (SWs), the collective spin angular momentum excitations in magnetically ordered material and their associated quanta, magnons, are considered the promising information carrier for the next-generation lower power consumption, higher speed, and higher density devices. The phase\cite{2005Khitun, 2014Klingler,2015Klingler,2017Fischer}, amplitude\cite{2005Kostylev, 2008Schneider, 2008Lee, 2013Jamali}, and polarization\cite{2017Lan, 2020Yu} of SWs can be employed to encode information. Generally, the nanostructured SW logic gates are designed on the uniform magnetized ferromagnetic materials by using all kinds of methods to control the propagation of SWs, including  magnetic field\cite{2013Jamali}, electric current\cite{2005Kostylev, 2008Schneider, 2008Lee} and voltage\cite{2017Wang, 2018Rana}.\par
The logic gates based on electric field or voltage can be easily programmable and compatible with nanoscale microwave devices\cite{2019Rana}. Rana et al.\cite{2018Rana} have proposed SW XNOR and universal NAND logic gates based on the voltage-controlled magnetic anisotropy (VCMA), which mainly arises from the change of electronic occupation state near the interface between the magnetic films and nonmagnetic heavy metals controlled by the voltage\cite{2008Duan, 2009Tsujikawa, 2019Rana}.
Magnetoelectric effects\cite{2005Nicola, 2015Taniyama, 2017Song} provide another path for electric-field control of SWs in multiferroic materials\cite{2018Balinskiy, 2018Sadovnikov}. The most representative multiferroic system is artificial multiferroic materials. Just recently, electric field controlled spin wave propagation had been experimentally demonstrated in artificial multiferroic heterostructure.\cite{2021Qin} Magnetoelectric effects could help us to design much more functional SW logic gates based on multiferroic structure. 
Here, the dual function SW logic gates are proposed by controlling the motion of magnetic anisotropy boundary (MAB) in epitaxial Fe film on ferroelectric $ \rm BaTiO_3$ substrate with altering in-plane and out-of-plane polarization domains. The AND-OR gate and NAND-NOR gate are validated by micromagnetic simulation. 

\begin{figure}
	\centering
	\includegraphics{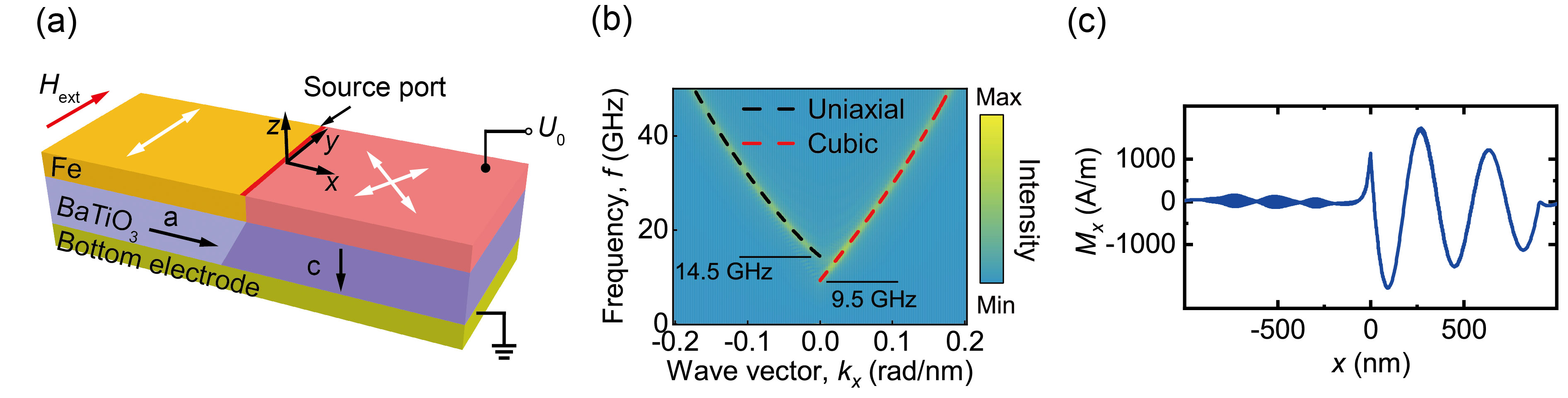}
	\caption{ (a) Schematic of the simulation model. Fe film is grown on the $ \rm BaTiO_3 $ with alternating in-plane polarization a-domain and out-of-plane polarization c-domain. The origin of the coordinate system is set at the center of the Fe film. 
	(b) The image plot of the dispersion relation of the SWs. The dash lines are theoretically calculated results. (c) 
	A snapshot of the $M_x$ component of the magnetization taken at \emph{t} = 4 ns, \emph{y} = $-1$ nm with a fixed \emph{f} = 12.6 GHz.}
	\label{figure1}
\end{figure}

We perform micromagnetic simulations by MuMax3\cite{2014Vansteenkiste} to study the propagation of Damon-Eshbach (DE) SWs\cite{1961Damon} in the epitaxial Fe film on ferroelectric $ \rm BaTiO_3$ substrate.
Figure \ref{figure1}(a) shows one unit structure in our simulation, including one 2 nm thick Fe layer, 100 nm thick $ \rm BaTiO_3$, and the bottom electrode layer. The width and length of the structure are 800 nm and 2000 nm, respectively.
The magnetic Fe layer includes uniaxial and cubic magnetic anisotropy regions via inverse magnetostriction owning to the different strains between Fe film and two kinds of domains in $ \rm BaTiO_3$ layer\cite{2012Lahtinen}. The easy axis of uniaxial magnetic anisotropy on the top of in-plane ferroelectric a-domain is along the \emph{y} axis, while the easy axes of cubic anisotropy on the top of out-of-plane ferroelectric c-domain are along the direction with 45\textdegree \ to the \emph{x} axis, as shown by the white double head arrows. The MAB is at the middle of the Fe film (\emph{x} = 0, along \emph{y} axis) and is pinned on the ferroelectric domain wall of the $\rm BaTiO_3$ substrate. 
The width of MAB in Fe film is same to the ferroelectric domain wall in $\rm BaTiO_3$ (2--5 nm\cite{1992Zhang, 2006Hlinka, 2006ZhangQingsong}), which is far smaller than the wavelength of SWs (several hundred nanometers). So, we neglect the width of the MAB in simulation. 
The parameters of Fe film used in simulation are the following\cite{2015Franke}: saturation magnetization $M_{\mathrm{s}}$ = 1.7 $\times$ $10^6$ A/m, exchange stiffness constant $A_{\mathrm{ex}}$ = 2.1 $\times$ $10^{-11}$ J/m, Gilbert damping constant $\alpha$ = 0.01. The uniaxial and cubic magnetic anisotropy constants are the experiment values\cite{2021Qin} of $K_{\mathrm{u}}$ = 1.5 $\times$ $10^4$ $\rm{J/m^3}$ and $K_{\mathrm{c}}$ = 4.4  $\times$ $10^4$ $\rm{J/m^3}$. The cell size is 2 $\times$ 2 $\times$ 2 $\rm nm^3$, which is smaller than the exchange length ($l_{\mathrm{ex}} \approx 3.4$ nm). In order to avoid SWs’ reflection at the boundary, the damping is set to 1 at both ends (width = 100 nm) of the Fe film. An external magnetic field $\mu _0 H_{\mathrm{ext}}$ = 100 mT is applied to avoid the formation of magnetic domain wall and magnetize the Fe film along \emph{y} axis. 
To obtain the dispersion relation of the DE SWs in uniaxial and cubic anisotropy regions, a sinc based exciting field, $\bm{\mathrm{h}}(t)=h_0\mathrm{sinc}(2\pi f_\mathrm{c}(t-t_0))\hat{e}_z$, with $\mu _0 h_0$ = 10 mT, $f_\mathrm{c}$ = 50 GHz, $t_0$ = 5 ns, is applied locally to a 2 $\times$ 800 $\times$ 2 $\rm nm^3$ central section of the Fe film, which is the SW source port (S) as indicated by the red region in \Fig{figure1}(a).
The dispersion relation\cite{2012Kumar,2013Venkat} of SWs can be obtained by performing a two-dimensional Fourier transform on $m_x$ ($m_x = M_x/M_\mathrm{s}$), as shown in \Fig{figure1}(b). The curve of the left branch is the dispersion relation of SWs propagating along the the negative direction of \emph{x} axis in the uniaxial anisotropy region and the right branch is the dispersion relation of SWs propagating along the positive direction of \emph{x} axis in the cubic anisotropy region. The theory curves of SW dispersion relations are calculated by adopting the Eq.\ref{fKu} and Eq.\ref{fKc}. 
It is clear to show that the SWs are only excited in the cubic anisotropy region in the frequency range of 9.5 GHz \textless \ \emph{f} \textless \ 14.5 GHz.
To further clarify the property, we excite SWs at the source port with a sinusoidal field $\bm{\mathrm{h}}(t) = h_0\sin (2\pi ft)\hat{e}_z$ with $\mu _0h_0$ = 10 mT and a fixed \emph{f} = 12.6 GHz. A snapshot of the $M_x$ component of the magnetization taken at \emph{t} = 4 ns is presented in \Fig{figure1}(c). 
SWs only propagate in the region with cubic anisotropy.\par
An out-of-plane electric field controlling ferroelectric domain wall motion can realize the movement of the MAB\cite{2015Franke}. When the electric field is along the direction of the polarization in the c-domain, the c-domain will expand while the a-domain shrinks by the lateral domain wall motion. Suppose the electric field is inverse to the direction of the polarization in the c-domain. In that case, the c-domain will shrink and the a-domain expands. So, the electric field can drive the ferroelectric domain wall's motion, thus driving the concurrent motion of the MAB. 
The velocity of MAB, that is, the velocity of ferroelectric domain wall driven by electric field pulse, can be up to 1000 m/s\cite{1963Stadler, 2017Boddu}.
Because the wall velocity is still smaller than the speed of sound in $\rm BaTiO_3$ ($v_{\mathrm{sound}}$ = 2000 m/s\cite{2010Tagantsev}), the strain state can be considered quasistatic during the motion of ferroelectric domain wall and the strength and symmetry of the magnetic anisotropy are invariant during the MAB back and forth.
The velocity $v$ of the MAB is implemented by shifting the MAB over one discretization cell ( $\delta x$ = 2 nm) during each time window $\delta t = \delta x/v$. In the simulation, the velocity of MAB is 1000 m/s under the electric field 31.4 MV/m obtained by applying 3.14 V perpendicular voltage, which is inferred from previous experimental results\cite{1963Stadler}.\par

\begin{figure}[!htb]
	\centering
    \includegraphics{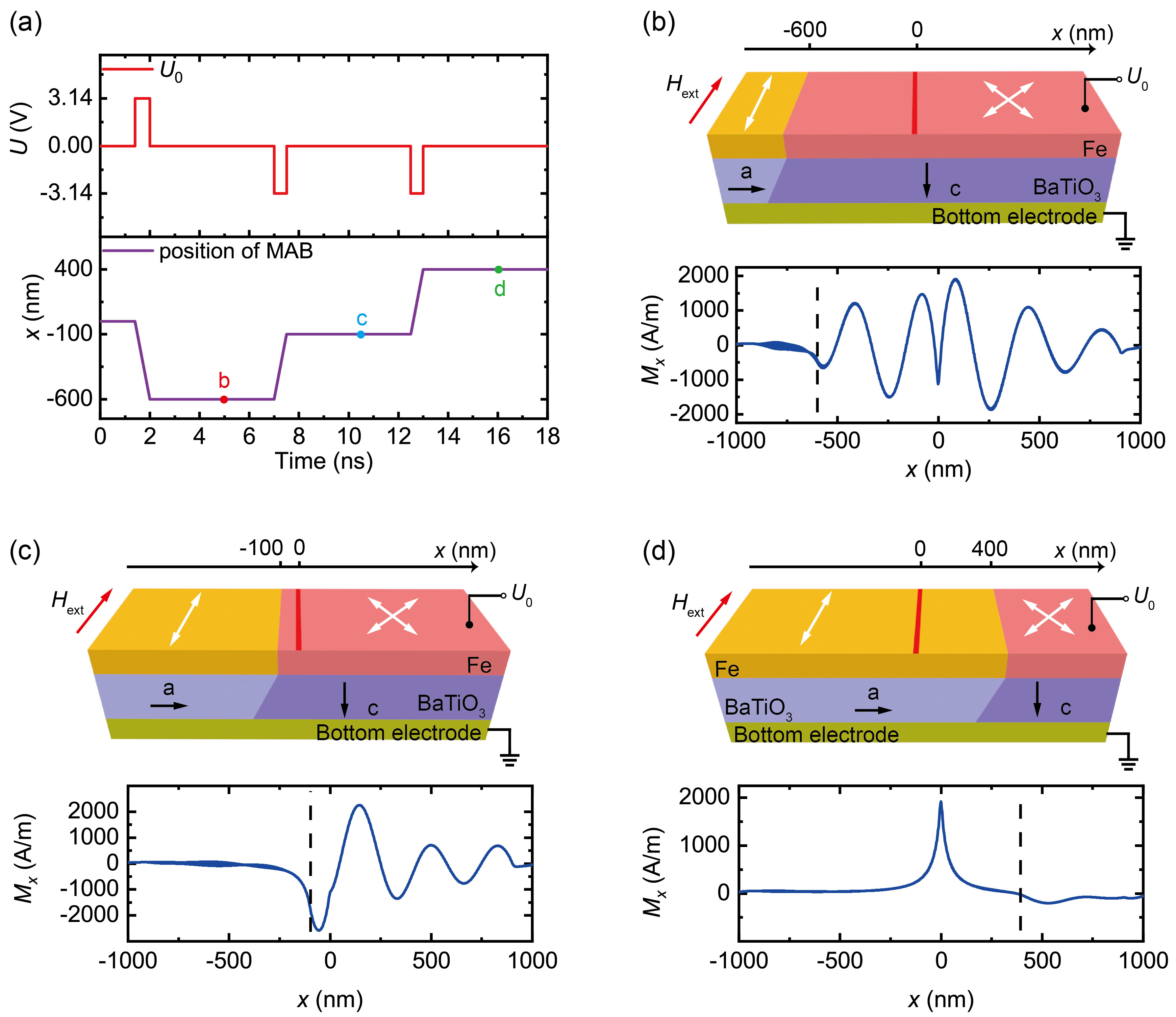}	
	\caption{ (a) The position variation of the MAB with voltage pulse. (b), (c) and (d) are the schematics of model and the snapshots of $M_x$ taken at \emph{y} = $-1$ nm when MAB at \emph{x} = $-600$ nm, $-100$ nm and 400 nm (adopting the coordinate system in \Fig{figure1}(a)), respectively. The dashed lines mark the position of MAB.
	}
	\label{figure2}
\end{figure}

Figure \ref{figure2}(a) shows the motion of MAB driven by voltage pulse and \Fig{figure2}(b)--(d) shows the schematics of model and the snapshots of $M_x$ when MAB at \emph{x} = $-600$ nm, $-100$ nm and 400 nm, respectively. The SWs with \emph{f} = 12.6 GHz are excited from the source port. The position of MAB is driven to \emph{x} = $-600$ nm by using the positive electric voltage (1.4 ns -- 2ns) shown in \Fig{figure2}(a). 
As expected, the SWs are blocked in uniaxial anisotropy region shown by the snapshot of SWs taken at \emph{t} = 5 ns in \Fig{figure2}(b). 
After applying one negative pulse voltage (7 ns -- 7.5 ns), the MAB moves to the position \emph{x} = $-100$ nm. And, the SWs only propagate in the cubic anisotropy region from the snapshot of $M_x$ taken at \emph{t} = 10.5 ns, shown in \Fig{figure2}(c). When another 0.5 ns negative voltage pulse is applied (12.5 ns -- 13 ns), the MAB moves to the position \emph{x} = $400$ nm, and the SW source port is out-of cubic anisotropy region. 
There are no SWs in Fe film. Thus, the transmission and block of SWs in Fe film can be controlled by voltage through driving the motion of MAB. Based on this unique property, we propose the design of dual function SW logic gates.\par
\begin{figure}[!htb]
	\centering
	\includegraphics{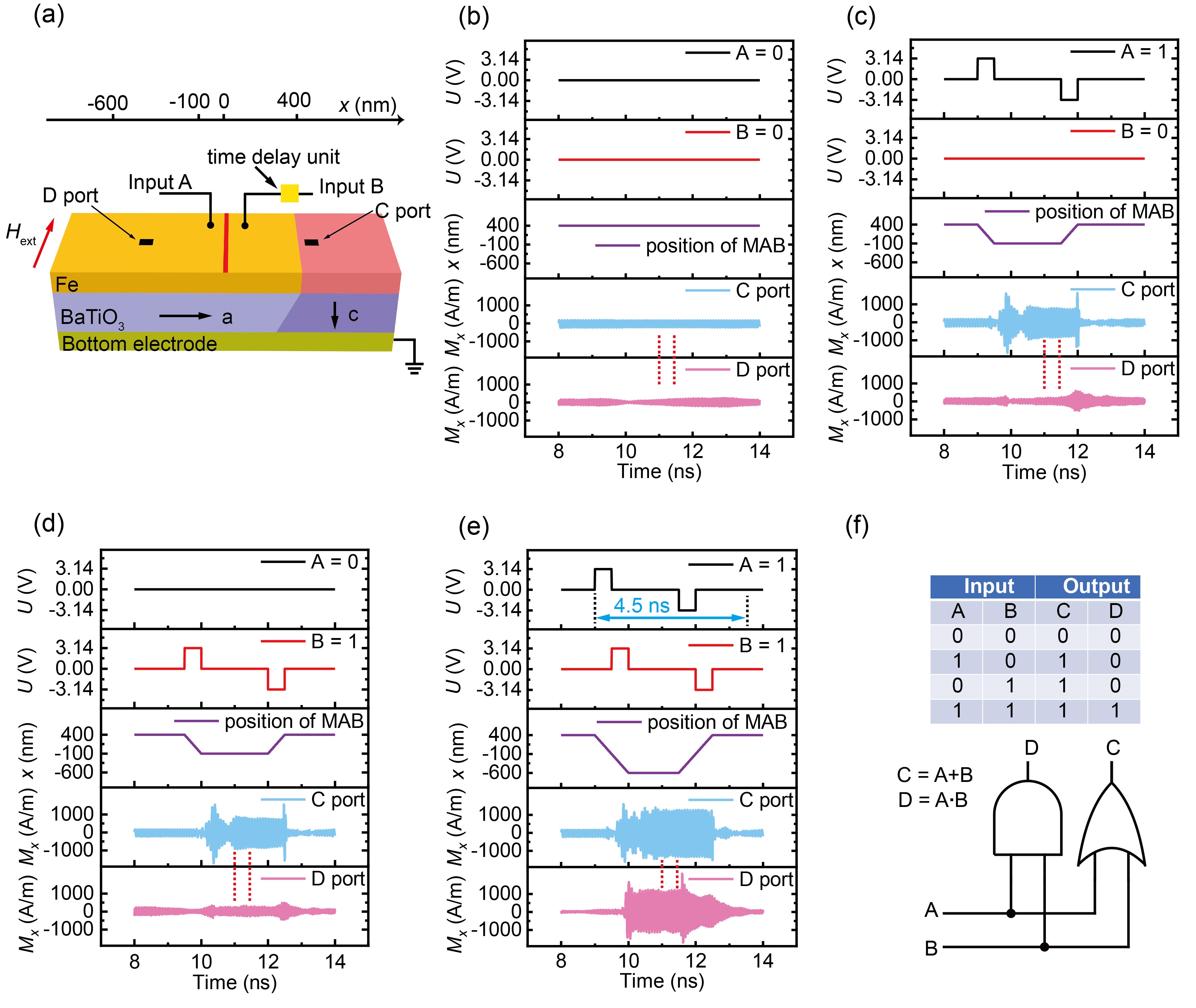}	
	\caption{ (a) Schematic diagram of the AND-OR gate. Initially, the MAB is set at \emph{x} = 400 nm. Input A and B are the input ports and input B includes a time delay unit that can delay the input signal for 0.5 ns. C and D ports are output ports. 
	The bottom electrode is grounding.
	The simulation results of the AND-OR gate with different inputs are shown in (b) -- (e), respectively. The amplitudes of SWs in 11 ns -- 11.5 ns are the logic operation results and the time domain is marked by the red dash lines. Finishing one logic operation needs 4.5 ns. 
	(f) The truth table of the AND-OR gate.
	}
	\label{figure3}
\end{figure}

Figure \ref{figure3}(a) is the schematic of the dual function AND-OR gate, which includes two input ports (A and B), two output ports (C and D), and a source port of SWs.
Initially, the MAB is set at \emph{x} = 400 nm parallel to \emph{y} axis. The position of input A and B ports are arbitrary on the top of the Fe film. In order to make the two input ports independent, the B port includes a time delay unit, which can delay the voltage pulse for 0.5 ns. 
The output ports C and D take the \emph{x} component of magnetization at region (498 nm \textless \ \emph{x} \textless \ 500 nm, $-2$ nm \textless \ \emph{y} \textless \ 0 nm) and ($-500$ nm \textless \ \emph{x} \textless \ $-498$ nm, $-2$ nm \textless \ \emph{y} \textless \ 0 nm), respectively, as indicated by the black square regions. Magnetic tunnel junctions can perform this function and convert SW signals to electrical signals. 
An external magnetic field $\mu _0 H_{\mathrm{ext}}$ = 100 mT is applied to magnetize the Fe film along \emph{y} axis and a sinusoidal field $\bm{\mathrm{h}}(t) = h_0 \sin(2\pi ft)\hat{e}_z$ with $\mu _0h_0$ = 10 mT and \emph{f} = 12.6 GHz is set at the source port. The bottom electrode is grounding.\par
The simulation results of the AND-OR gate with different inputs are shown in \Fig{figure3}(b)--(e). 
The simulation result of A = 0 and B = 0 is shown in \Fig{figure3}(b). 
Obviously, the MAB keeps still at position \emph{x} = 400 nm, because there is no electric field across $\rm BaTiO_3$. The outputs of C and D are both small amplitudes representing '0' due to the blocking of SWs shown in \Fig{figure2}(d). 
The simulation result of A = 1 and B = 0 is shown in \Fig{figure3}(c). The input '1' signal includes one 0.5 ns positive voltage pulse and 0.5 ns negative voltage pulse as shown in \Fig{figure3}(c). The positive pulse is used to start logical operation by driving the MAB moving 500 nm along the negative direction of \emph{x} axis in 0.5 ns, while the negative pulse is used to reset the position of MAB.
The related motion of MAB and output signals of the two ports are also shown in \Fig{figure3}(c).
In 9 ns -- 9.5 ns, the MAB is driven from \emph{x} = 400 nm to \emph{x} = $-100$ nm. After oscillation in 9.5 ns -- 11 ns, 
the amplitudes of SWs at C and D represent C = 1 and D = 0 in the time range of 11 ns -- 11.5 ns, marked by the red dash lines. After that, the negative pulse drives the MAB back to \emph{x} = 400 nm during 11.5 ns -- 12 ns. The outputs will be the same as the initial and one whole logical operation is finished.
\Fig{figure3}(d) shows the simulation result of A = 0 and B = 1. It's very similar to the previous operation. The only difference is the 0.5 ns delay of the voltage pulse in B port. But the outputs of ports C and D are still '1' and '0', respectively. 
If A and B ports are both '1', the driving time of MAB will be 1 ns due to the 0.5 ns delay between A and B shown in \Fig{figure3}(e). The MAB is driven to \emph{x} = $-600$ nm from \emph{x} = 400 nm during the period of 9 ns -- 10 ns. 
And the amplitudes of SWs for ports C and D are over 1000 A/m at the time range of 11 ns -- 11.5 ns. The operation results of C and D are both '1'. Then, the MAB is driven back to \emph{x} = 400 nm during 11.5 ns -- 12.5 ns. 
It should be mentioned that the amplitude of the SWs in C port at 11 ns -- 11.5 ns is a bit larger than that in \Fig{figure3}(c) and (d), which mainly arises from the weaker interference between the reflected SWs from the MAB and the initial SWs. In short summary, the logic operation of the AND-OR gate can be divided into three stages, preparation before output (9 ns -- 11 ns), output (11 ns -- 11.5 ns) and reset (11.5 ns -- 13.5 ns). 
The truth table of the AND-OR gate is shown in \Fig{figure3}(f). The input A, B and output C compose an OR gate while the input A, B and output D compose an AND gate. \par 
\begin{figure}[!htb]
	\centering
	\includegraphics{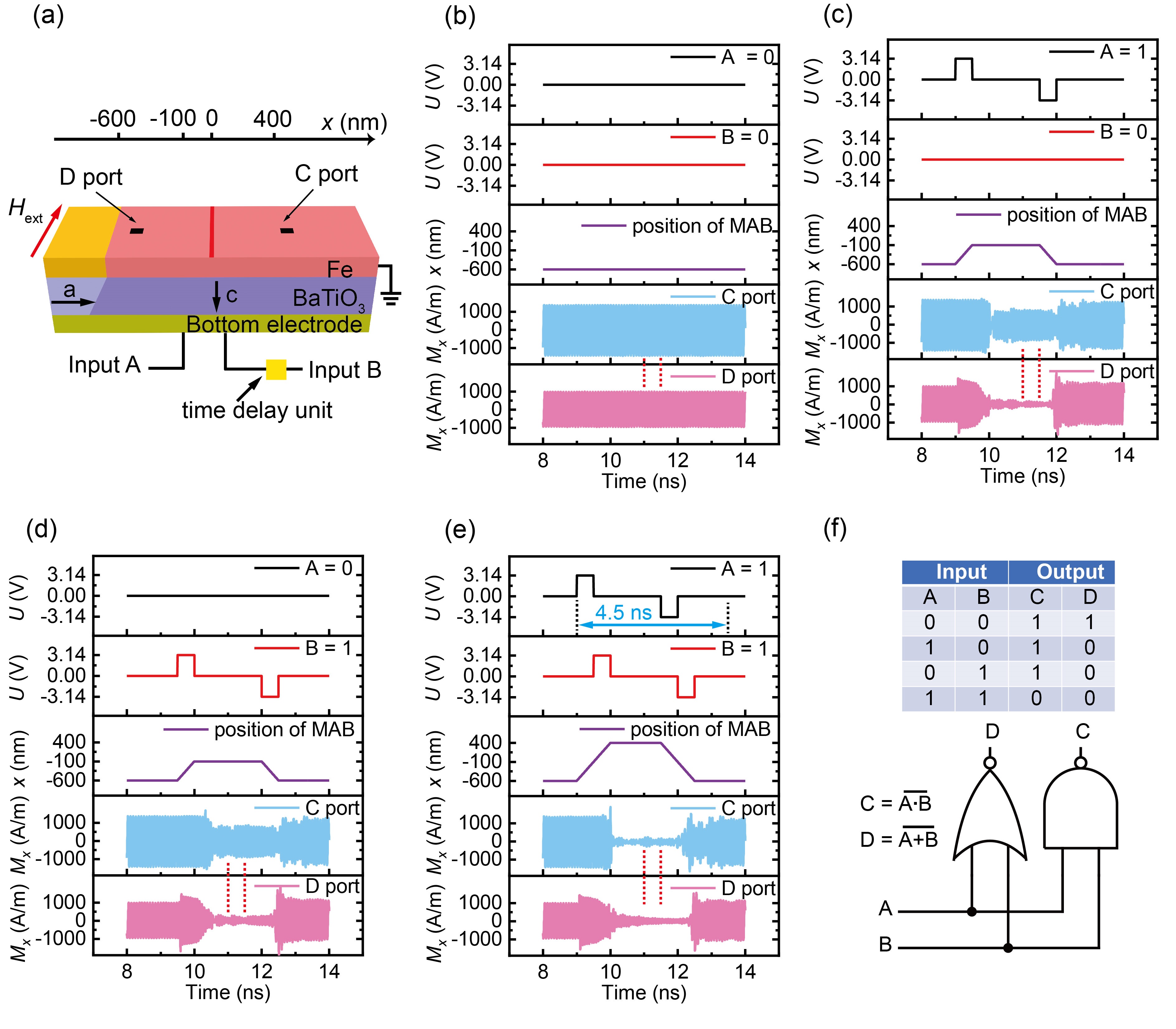}	
	\caption{ (a) Schematic diagram of the NAND-NOR gate. Initially, the MAB is set at \emph{x} = $-600$ nm. The Fe film is grounding and the input voltage signals are injected from bottom electrode. Other configurations and the input signals are same to the AND-OR gate.
	The simulation results of the NAND-NOR gate with different inputs are shown in (b) -- (e), respectively. The amplitudes of SWs in 11 ns -- 11.5 ns are the logic operation results and the time domain is marked by the red dash lines. Finishing one logic operation needs 4.5 ns. (f) The truth table of the NAND-NOR gate.
	}
	\label{figure4}
\end{figure}
In order to implement full Boolean logic operation, we also need the NAND and NOR gates. \Fig{figure4}(a) is the schematic of the dual function NAND-NOR gate. Here, the initial MAB is set at \emph{x} = $-600$ nm.  The main change of the NAND-NOR gate is that the two input ports A and B  are connected with the bottom electrode and the Fe layer is grounding. The output ports and source port are the same with the AND-OR gate.  
When A and B are both '0', the MAB keeps still at \emph{x} = $-600$ nm. When only one of A and B is '1', the MAB is driven to \emph{x} = $-100$ nm for logic operation and then driven back to reset. When A and B are both '1', the MAB is driven to \emph{x} = 400 nm for logic operation and then driven back to reset.
The simulation results of the NAND-NOR logic gate with different inputs are shown in \Fig{figure4}(b)--(e), respectively. 
The output result of A = 0 and B = 0 is shown in \Fig{figure4}(b). The MAB is still at \emph{x} = $-600$ nm because of the zero-field force. The amplitudes of the SWs are over 1000 A/m, both for C and D ports.
This indicates that the output ports C and D are both '1'.
The simulation result of A = 1 and B = 0 is shown in \Fig{figure4}(c). 
The MAB is driven from \emph{x} = $-600$ nm to \emph{x} = $-100$ nm for positive pulse during 9 ns -- 9.5 ns. Then, the amplitude of the SWs at C port is slightly reduced before the negative reset pulse, which can be explained by the interference between the reflected SWs from MAB and the initial SWs. For the D port, the amplitude of the SWs is almost vanishing. So we can recognize the output C = 1 and D = 0 before the reset of the MAB. 
\Fig{figure4}(d) shows the simulation result for A = 0 and B = 1. The output is same as A = 1 and B = 0.
The last configuration of A = 1 and B = 1 is shown in \Fig{figure4}(e). 
In 9 ns -- 10 ns, the MAB is driven to \emph{x} = 400 nm from \emph{x} = $-600$ nm. After oscillation in 10 ns -- 11 ns, the amplitudes of C and D in 11 ns -- 11.5 ns are the operation result representing C = 0 and D = 0. In 11.5 ns -- 12.5 ns, the MAB is driven back to \emph{x} = $-600$ nm. After oscillation in 12.5 ns -- 13.5 ns, one logic operation is finished. 
The logic operation of the NAND-NOR gate can also be divided into three stages, preparation before output (9 ns -- 11 ns), output (11 ns -- 11.5 ns) and reset (11.5ns -- 13.5 ns). One circle logic operation needs 4.5 ns. 
The truth table of the NAND-NOR gate is listed in \Fig{figure4}(f). The input A, B and output C compose a NAND gate, while the input A, B and output D compose a NOR gate. \par 
In conclusion, we propose dual function SW logic gates based on the motion of MAB controlled by electric field. AND-OR gate and NAND-NOR gate are validated by micromagnetic simulation. 
Owning to the inherent dual function, the proposed SW logic gates have potential of building low power consumption and high density devices. Our results will also motivate further experimental studies for the development of dual function SW devices.
\begin{acknowledgments}
This work is partially supported by National Key Research
Program of China (Grant No. 2017YFA0206202), International Postdoctoral Exchange Fellowship Program (20190083), Natural Science Foundation of Shaanxi Province (2021JM-022).
\end{acknowledgments}

\appendix
\section{Dispersion relation of SWs}
Kalinikos et al\cite{1986Kalinikos} developed the theory of dispersion relation of SWs taking into account both dipole-dipole and exchange interactions. The general formula of dispersion relation of DE SWs is :
\begin{equation}
    f = \frac{\gamma \mu _0}{2\pi}\sqrt{(H_i + \frac{2A_{\mathrm{ex}}}{\mu_0 M_{\mathrm{s}}}k^2)(H_i + M_{\mathrm{s}} + \frac{2A_{\mathrm{ex}}}{\mu_0 M_{\mathrm{s}}}k^2) + M^2_{\mathrm{s}}F}
    \label{DE_general}
\end{equation}
where $\gamma$ is the gyromagnetic ratio, $\mu _0$ is vacuum permeability, $H_i$ is the internal static magnetic field, $A_{\mathrm{ex}}$ is exchange stiffness constant, $M_{\mathrm{s}}$ is saturation magnetization, $k$ is the wave vector of SWs, $F = \frac{1-\mathrm{e}^{-kd}}{kd}(1-\frac{1-\mathrm{e}^{-kd}}{kd})$, $d$ is the thickness of the film. This formula doesn't include magnetic anisotropy field. So we need calculate magnetic anisotropy field and give the formula of dispersion relation in uniaxial and cubic anisotropy region, respectively.\par
The dynamic unit magnetization can be written as:
$\mathbf{m} = \frac{\mathbf{M}}{M_\mathrm{s}} \approx \hat{e}_y + m_x\hat{e}_x + m_z\hat{e}_z$,
where $m_x$, $m_z$ $\ll$ 1.
The magnetic anisotropy fields in uniaxial and cubic anisotropy regions can be calculated by equation (\ref{Ku}) and (\ref{Kc}) (ignore high-order $m_x$ and $m_z$ terms for linear approximation).
\begin{equation}
    \mathbf{H}_{\mathrm{u}} = \frac{2K_\mathrm{u}}{\mu _0 M_\mathrm{s}}(\mathbf{u}\cdot \mathbf{m})\mathbf{u} = \frac{2K_\mathrm{u}}{\mu _0 M_\mathrm{s}}\hat{e}_y
    \label{Ku}
\end{equation}
where $\mathbf{H}_\mathrm{u}$ is the uniaxial  anisotropy field, $K_\mathrm{u}$ is the uniaxial anisotropy constant, $\mathbf{u}$ is the unit vector of the easy axis of uniaxial anisotropy, along \emph{y} axis.
\begin{equation}
    \begin{aligned}
        \mathbf{H}_\mathrm{c} = -\frac{2K_\mathrm{c}}{\mu _0 M_\mathrm{s}}(\quad &((\mathbf{c_2} \cdot \mathbf{m})^2 + (\mathbf{c_3} \cdot \mathbf{m})^2)((\mathbf{c_1} \cdot \mathbf{m})\mathbf{c_1}) \quad + \\
        &((\mathbf{c_1} \cdot \mathbf{m})^2 + (\mathbf{c_3} \cdot \mathbf{m})^2)((\mathbf{c_2} \cdot \mathbf{m})\mathbf{c_2}) \quad + \\
        &((\mathbf{c_1} \cdot \mathbf{m})^2 + (\mathbf{c_2} \cdot \mathbf{m})^2)((\mathbf{c_3} \cdot \mathbf{m})\mathbf{c_3}) \quad ) \\
        &=\frac{K_\mathrm{c}}{\mu _0M_\mathrm{s}}m_x\hat{e}_x - \frac{K_\mathrm{c}}{\mu _0M_\mathrm{s}}\hat{e}_y - \frac{2K_\mathrm{c}}{\mu _0M_\mathrm{s}}m_z\hat{e}_z
    \end{aligned}
    \label{Kc}
\end{equation}
where $\mathbf{H}_\mathrm{c}$ is the cubic anisotropy field, $K_\mathrm{c}$ is the cubic anisotropy constant, $\mathbf{c}_1$, $\mathbf{c}_2$ and $\mathbf{c}_3$ are ($\sqrt{2}/2$, $\sqrt{2}/2$, 0), ($-\sqrt{2}/2$, $\sqrt{2}/2$, 0) and (0, 0, 1), respectively, the unit vector of the easy axes of cubic anisotropy. \par

So, the formulas of dispersion relation of DE SWs in uniaxial and cubic anisotropy region are 
\begin{equation}
    f = \frac{\gamma \mu _0}{2\pi}\sqrt{(H_{\mathrm{ext}}+\frac{2K_{\mathrm{u}}}{\mu _0M_{\mathrm{s}}} + \frac{2A_{\mathrm{ex}}}{\mu_0 M_{\mathrm{s}}}k^2)(H_{\mathrm{ext}}+\frac{2K_{\mathrm{u}}}{\mu _0M_{\mathrm{s}}} + M_{\mathrm{s}} + \frac{2A_{\mathrm{ex}}}{\mu_0 M_{\mathrm{s}}}k^2) + M^2_{\mathrm{s}}F}
      \label{fKu}
\end{equation} and
\begin{equation}
    f = \frac{\gamma \mu _0}{2\pi}\sqrt{(H_{\mathrm{ext}}-\frac{2K_{\mathrm{c}}}{\mu _0M_{\mathrm{s}}} + \frac{2A_{\mathrm{ex}}}{\mu_0 M_{\mathrm{s}}}k^2)(H_{\mathrm{ext}}+\frac{K_{\mathrm{c}}}{\mu _0M_{\mathrm{s}}} + M_{\mathrm{s}} + \frac{2A_{\mathrm{ex}}}{\mu_0 M_{\mathrm{s}}}k^2) + M^2_{\mathrm{s}}F}
    \label{fKc}
\end{equation}

\bibliographystyle{rsc} 
\bibliography{Bib}

\end{document}